\documentclass[preprint,aps,showpacs,nofootinbib,preprintnumbers,amsmath,amssymb,superscriptaddress]{revtex4-2}
\usepackage{amsmath,amssymb,amsthm,mathrsfs}
\usepackage{epsfig}
\usepackage{graphicx}
\usepackage[usenames,dvipsnames]{color}
\usepackage{subfigure}
\usepackage{slashed}
\usepackage{comment}
\usepackage[normalem]{ulem}
\usepackage{tikz}
\usepackage[compat=1.1.0]{tikz-feynman}
\usepackage[colorlinks,citecolor=blue]{hyperref}
\usepackage{color}

\usepackage{multirow}
\usepackage{tikz-feynman}
\tikzfeynmanset{compat=1.1.0}

\usepackage{tikzsymbols}

\usepackage{xcolor}
\usepackage{soul}

\begin{document}
\title{Gravitational Wave Birefringence from Fuzzy Dark Matter}
\author{Da~Huang}
\email{dahuang@bao.ac.cn}
\affiliation{National Astronomical Observatories, Chinese Academy of Sciences, Beijing, 100012, China}
\affiliation{School of Fundamental Physics and Mathematical Sciences, Hangzhou Institute for Advanced Study, UCAS, Hangzhou 310024, China}
\affiliation{University of Chinese Academy of Sciences (UCAS), Beijing 100049, China}

\author{Ze-Xuan Xiong}
\email{xiongzexuan22@mails.ucas.ac.cn}
\affiliation{School of Fundamental Physics and Mathematical Sciences, Hangzhou Institute for Advanced Study, UCAS, Hangzhou 310024, China}
\affiliation{University of Chinese Academy of Sciences (UCAS), Beijing 100049, China}
\affiliation{Institute of Theoretical Physics, Chinese Academy of Sciences, Beijing 100190, China}

\author{Lei-Jian Wang}
\email{wangleijian24@mails.ucas.ac.cn}
\affiliation{School of Fundamental Physics and Mathematical Sciences, Hangzhou Institute for Advanced Study, UCAS, Hangzhou 310024, China}
\affiliation{University of Chinese Academy of Sciences (UCAS), Beijing 100049, China}
\affiliation{Institute of Theoretical Physics, Chinese Academy of Sciences, Beijing 100190, China}

\date{\today}
\begin{abstract}
\noindent Gravitational wave (GW) birefringence is a remarkable phenomenon that can be used to test the parity violation in gravity. By coupling the fuzzy dark matter (FDM) scalar to the gravitational Chern-Simons term, we explore the GW birefringence effects in the FDM background. In particular, in light of the highly oscillating granular FDM structure at the galactic scale, we are led to investigating the GW propagation in the Chern-Simons gravity over the general nontrivial scalar profile, which is a natural extension of previous studies on the homogeneous and isotropic configurations. As a result, it is found that GWs of both circularly polarized modes propagate in the straight line with the speed of light, and does not show any velocity birefringence. However, when considering the imaginary part of the dispersion relation, GWs exhibit the amplitude birefringence in which one circular polarization is enhanced while the other suppressed. Due to its local nature, the FDM-induced amplitude birefringence factor only depends on the GW frequency without any reliance on the GW propagating distance, which can be used to distinguish this signal from other birefringece mechanisms. More importantly, the birefringence shows a periodic time modulation with the period directly reflecting the FDM scalar mass, which is another smoking gun for testing this model. Finally, we also study the extra-galactic FDM contribution to the GW birefringence, which is shown to be suppressed by the cosmological DM density and thus subdominant compared with the galactic counterpart. 

\end{abstract}

\maketitle
\newpage

\section{Introduction}\label{s1}
Testing gravitational parity violation is a key to understand the nature of gravity. It is well-known that general relativity (GR) preserves the parity symmetry. However, the parity conservation can be broken in many modifications to GR~\cite{Nojiri:2017ncd}, such as Chern-Simons (CS) gravity~\cite{Lue:1998mq,Jackiw:2003pm,Alexander:2009tp}, Teleparallel Gravity~\cite{Conroy:2019ibo,Crisostomi:2017ugk}, ghost-free scalar-tensor gravity~\cite{Nishizawa:2018srh} and so on~\cite{Hojman:1980kv,Horava:2009uw,Takahashi:2009wc,Zhu:2013fja,Altschul:2009ae,Takahashi:2022mew,Sulantay:2022sag,Bombacigno:2022naf,Boudet:2022nub,Kawai:2017kqt,Kostelecky:2016kfm,Kanno:2023kdi}. Moreover, the direct observation of gravitational waves (GWs) by the LIGO-Virgo-Kagra (LVK) collaboration~\cite{LIGOScientific:2016aoc,LIGOScientific:2017zic,LIGOScientific:2018mvr,LIGOScientific:2020ibl,KAGRA:2021vkt} has opened a new avenue to probe this important issue. Especially, one smoking gun of parity violation in gravity is provided by the GW birefringence effects~\cite{Alexander:2004wk,Alexander:2007kv,Yunes:2010yf,Alexander:2017jmt,Ezquiaga:2021ler}, {\it i.e.}, the two circularly polarized modes behave differently in phase and amplitude when propagating over astrophysical and cosmological distances. Such a remarkable phenomenon has already been widely studied in many modified gravity theories~\cite{Satoh:2007gn,Kato:2015bye,Yoshida:2017cjl,Jung:2020aem,Chu:2020iil,Li:2020xjt,Okounkova:2021xjv,Nojiri:2019nar,Daniel:2024lev,Manton:2024hyc,Lagos:2024boe,Alexander:2024klf,Alexander:2025wnj,Horii:2025jen} and in model-independent ways~\cite{Ng:2023jjt,Zhao:2019xmm,Wang:2020cub,Jenks:2023pmk,Callister:2023tws,Guo:2025bxz,Jia:2026vqo}.

Although more and more astrophysical evidences for dark matter (DM) have been accumulated in the past several decades~\cite{Bertone:2016nfn}, its nature is still elusive. More recently, the fuzzy dark matter (FDM)~\cite{Hu:2000ke,Marsh:2010wq,Hui:2016ltb,Berezhiani:1992rk} has become a promising DM candidate since it may provide possible solutions to many problems observed on sub-galactic scales (see {\it e.g.} Refs.~\cite{Niemeyer:2019aqm,Ferreira:2020fam,Hui:2021tkt} for recent reviews). Importantly, latest precise N-body simulations~\cite{Schive:2014dra,Schive:2014hza,Mocz:2017wlg,Veltmaat:2016rxo,Nori:2018hud} have shown that, for a FDM with its mass of ${\cal O}(10^{-22}\,{\rm eV})$, the FDM scalar field is oscillating in time and forms a solitonic core surrounded by DM halos with complicated granular structures. Moreover, it was argued that the FDM can be identified as an axion-like particle~\cite{Khlopov:1998uj,Hu:2000ke,Marsh:2010wq,Hui:2016ltb,Berezhiani:1992rk,Niemeyer:2019aqm,Hwang:2021vuq,Qiu:2022uvt} since its lightness can be easily understood by its associated approximate shift symmetry. Thus, it is quite natural to expect that the FDM can have the gravitational CS coupling arising either from the gravitational anomaly~\cite{Alvarez-Gaume:1983ihn,Weinberg:1996kr} or in the string theory~\cite{Green:1987mn,Polchinski:1998rr,Alexander:2004us,Alexander:2004xd}. Therefore, all these FDM properties motivate us to consider the GW birefringence phenomena in the CS gravity when GWs propagate in a general FDM profile of highly non-trivial spacetime dependence, which can be viewed as an extension of the previous framework~\cite{Alexander:2007kv,Yunes:2010yf,Alexander:2017jmt,Ezquiaga:2021ler,Satoh:2007gn,Kato:2015bye,Yoshida:2017cjl,Li:2020xjt,Jung:2020aem,Okounkova:2021xjv,Nojiri:2019nar,Daniel:2024lev,Manton:2024hyc,Lagos:2024boe,Zhao:2019xmm,Wang:2020cub,Jenks:2023pmk,Ng:2023jjt} by assuming a homogeneous and isotropic scalar profile. Given that the FDM background variations in time and space are much smaller than the GW frequency and wavenumber, we shall perform our calculation with the well-known eikonal approximation~\cite{weinberg1962eikonal}.  As a result, we can show that the FDM-induced GW birefringence exhibits many new features, which can be used to distinguish this signal from other cosmologically generated ones. Finally, we further consider the extra-galactic FDM contribution to the GW birefringence and assess its impact on the detectability of the galactic one. Here we would like to mention that the GW propagation in the non-trivial axion DM background, especially in the Milky Way (MW), has recently been explored in the literature~\cite{Yoshida:2017cjl,Jung:2020aem,Chu:2020iil,Tsutsui:2022zos,Lambiase:2022ucu,Tsutsui:2023jbk}. However, these studies focused on the GW production via the axion resonant decays, in which the axion mass was assumed to be of ${\cal O}(10^{-10}\sim 10^{-12})~{\rm eV}$ in order to fit the GW energies of LVK events~\cite{Yoshida:2017cjl,Jung:2020aem,Chu:2020iil,Tsutsui:2022zos,Tsutsui:2023jbk}. Note that the GW signals generated in this mechanism are parity-conserving. In contrast, here we shall consider a very light FDM axion, with its mass around or smaller than $10^{-22}$~eV, which can lead to true parity-violating GW birefringence phenomena.  

The paper is structured as follows. We begin our discussion in Sec.~\ref{CSGraivty} by setting up our conventions of the CS gravity theory where the scalar field is identified as the FDM candidate. Then we present in Sec.~\ref{GWBiGen} the general formalism for the GW propagation over a coherent FDM background. In Sec.~\ref{GWBiGal}, we focus on the GW birefringence taking place inside our galaxy, and identify its observational signatures. Sec.~\ref{GWBiEx} is devoted to investigating the contribution to the GW birefringence from the extra-galactic FDM. Finally, we conclude and provide further discussions in Sec.~\ref{conclusion}.

\section{Chern-Simons Gravity Theory}\label{CSGraivty}
The GW birefringence can be generated if the FDM pseudoscalar $\phi$ possesses the gravitational CS coupling. Following the conventions in Ref.~\cite{Alexander:2009tp}, the Lagrangian of the CS gravity can be written as follows
\begin{eqnarray}
	{\cal S} = \int d^4x \sqrt{-g} \left[\kappa  R + \frac{\alpha}{4} \phi  ~^*R^{\tau}_{~\lambda \mu\nu} R^{\lambda~\mu\nu}_{~\tau}  \right]\,,
\end{eqnarray}
where $\kappa\equiv(16\pi G)^{-1}$ with $G$ the Newton constant, while $\alpha$ denotes the CS coupling with unit length dimension. The quantity $^*R^\tau_{~\lambda \mu\nu} R^{\lambda~\mu\nu}_{~\tau}$ is the so-called Pontryagin denisty. Here the dual Riemann tensor is defined by $^*R^{\lambda~\mu\nu}_{~\tau} \equiv \epsilon^{\mu\nu\rho\sigma} R^\lambda_{~\tau\rho\sigma}/2$, where ${\epsilon}^{\mu\nu\rho\sigma} = \tilde{\epsilon}^{\mu\nu\rho\sigma}/\sqrt{-g}$ is the 4-dimensional Levi-Civita tensor with the anti-symmetric symbol as $\tilde{\epsilon}^{0123} = -\tilde{\epsilon}_{0123}  = 1$. By differentiating the CS gravity action with respect to the tensor perturbation $h_{ij}$ defined by $ds^2 = a(\eta)^2 [-d\eta^2+( \delta_{ij}+h_{ij})dx^i dx^j]$,  we can obtain the following linearized gravitational equations~\cite{Alexander:2004wk}
\begin{eqnarray}\label{EoMGWt}
	\square h^j_{~i} =  \frac{\alpha}{\kappa} \epsilon^{pjk} \partial^\alpha \left[\frac{1}{a^2} \left( \partial_p \phi \partial_\eta \partial_\alpha - \partial_\eta \phi \partial_p  \partial_\alpha \right) \right]h_{ki}\,,
\end{eqnarray}
where $\square \equiv (-\partial_\eta^2-2{\cal H}\partial_\eta + \partial_i^2)/a^2$ is the four-dimensional d'Alembertian operator in the Friedmann-Lema\^{\i}tre-Robertson-Walker metric with ${\cal H}$ the conformal Hubble parameter. Throughout this article, the Latin letters refer to spatial indices while the Greek letters to spacetime ones. Also, we have simplified our formulae by using the usual transverse-traceless gauge conditions $\delta^{ij}h_{ij} = \partial^i h_{ij} = 0$. In our work, we assume that the GW production is the same as in GR, while the FDM affects the subsequent propagation of GWs. Depending on the realistic FDM profiles inside and outside of the MW, we can further simplify the above general GW equations.

\section{Gravitational Wave Birefringence in the Milky Way}\label{GWBiGal}
GW birefringence has been widely studied in the literature, but earlier explorations usually assumed a homogeneous scalar background in the CS gravity. Given the complicated FDM distributions in our galaxy, it is generally expected that the induced GW birefringence would exhibit several new features. 
\subsection{General Formalism}\label{GWBiGen}
Let us firstly focus on the GW propagation over a coherent inhomogeneous FDM field around our MW, so that the cosmic expansion can be ignored. By setting the scale factor to be $a=1$ and replacing the conformal time $\eta$ with the physical one $t$, 
we can simplify the general equation in Eq.~\eqref{EoMGWt} as follows
\begin{eqnarray}\label{EoMmin}
	 \square h^j_{~i} =  (\alpha/\kappa) \epsilon^{pjk} \partial^\alpha \left( \partial_p \phi \partial_t \partial_\alpha - \partial_t \phi \partial_p  \partial_\alpha \right) h_{ki}\,.
\end{eqnarray}
In order to proceed, we shall work in the coordinates where the GW moves in the $z$-direction, so that the widely-used linearly polarizations can be identified as $h_+ \equiv h_{xx} = -h_{yy}$ and $h_\times \equiv  h_{xy} = h_{yx}$. By defining the circularly polarized GW modes as~\cite{Isi:2022mbx}
\begin{eqnarray}
	h_{\rm R,L} = ({1}/{\sqrt{2}}) (h_+ \mp ih_\times)\,,
\end{eqnarray}
Eq.~\eqref{EoMmin} can be further decomposed into two equations for left- and right-handed polarizations
\begin{eqnarray}
	\square h_{\rm R,L} \mp i(\alpha/\kappa) \partial^\alpha \left[ \partial_z \phi \partial_\alpha \partial_t  - \partial_t \phi \partial_\alpha \partial_z \right] h_{\rm R,L} = 0\,.
\end{eqnarray}
where and in what follows the upper (lower) sign refers to the right(left)-handed mode. 


By further assuming that the time and spatial variations of the FDM scalar background are small compared with the typical GW frequency and wavenumber, we can take advantage of the famous eikonal approximation~\cite{weinberg1962eikonal,Blas:2019qqp} to compute the GW waveform corrections induced by the FDM. Concretely, we consider the GW solution of the form $h_{\rm R,L} = h^{0}_{\rm R,L} e^{iS}$, where $h_{\rm R,L}^{0}$ are slowly-varying amplitudes for both polarizations while the phase $S$ is fast evolving. 
According to the general rules of the eikonal approximation, we can define the following GW frequency and wavenumber
\begin{eqnarray}
	\omega = -\partial S/\partial t \,,\quad k_i=\partial S/\partial x^i\,.
\end{eqnarray} 
Since the GW propagates along $z$-axes by assumption, only $k_z=k$ is nonzero. The dispersion relations for both polarizations are given by
\begin{eqnarray}\label{GWdispersion}
	D^\pm = (\omega^2 - k^2) \left[1\mp \frac{\alpha}{\kappa} (\omega \partial_z \phi + k\partial_t \phi) \right]  \mp \frac{i \alpha}{\kappa} \left[ (\omega^2 + k^2) \partial_t \partial_z \phi + \omega k(\partial_z^2 \phi + \partial_t^2 \phi) \right] 
	= 0\,.
\end{eqnarray} 
Given the presence of the imaginary part in Eq.~\eqref{GWdispersion}, the GW frequency and wavenumber would, in general, be complex. In the literature, the real (imaginary) correction to the GW dispersion relation is often referred to the velocity (amplitude) birefringence, as it usually leads to different modifications to the speed (amplitude) for left- and right-handed GWs. It is interesting to note that the imaginary part in Eq.~\eqref{GWdispersion} is one-order smaller than the real correction in the limit $\partial_{t,z} \phi / \omega,\, k \ll 1$, so that we can consider them separately. 

\subsubsection{Velocity Birefringence}
We now take into account the real part of the dispersion relations in Eq.~\eqref{GWdispersion}
\begin{eqnarray}\label{GWd0}
	D_R^\pm = (\omega^2 - k^2) \left[1 \mp  (\alpha/\kappa)(\omega \partial_z \phi + k\partial_t \phi) \right] = 0\,,
\end{eqnarray}
which determines the GW velocity and direction in the FDM field $\phi$. It is clear that $\omega =k$ is always a solution to Eq.~\eqref{GWd0}\footnote{It seems that there is another solution to Eq.~\eqref{GWd0} with $\omega= [-k\partial_t\phi \mp \kappa/\alpha]/(\partial_z\phi)$. But this solution cannot continuously approach to the vacuum one $\omega=k$ and should be ignored. }, which indicates that GWs of both circular polarizations propagate with the speed of light, regardless of the presence of the FDM background. We can further check this by calculating GW paths under the influence of a nontrivial $\phi$ profile
\begin{eqnarray}\label{EoMGW0}
	\frac{dx^i}{dt}&=&-\frac{\partial D_R^\pm/\partial k_i}{\partial D_R^\pm/\partial \omega}  = \delta^i_z\,, \nonumber\\
	\frac{dk_i}{dt} &=& \frac{\partial D_R^\pm/\partial x^i}{\partial D_R^\pm/\partial \omega}=0\,,\quad
	\frac{d\omega}{dt} = -\frac{\partial D_R^\pm/\partial t}{\partial D_R^\pm/\partial \omega} = 0\,,
\end{eqnarray}
where we have used $\omega=k$ to simplify the these formulae. 
It is clear from Eq.~\eqref{EoMGW0} that both GW polarizations still follow straight lines, and their frequencies and wavenumbers remain the same during its galactic propagation. In a word, the FDM halo does not induce any velocity birefringence. 

\subsubsection{Amplitude Birefringence}
We shall turn to the full dispersion relation in Eq.~\eqref{GWdispersion} by including the imaginary part, which might lead to the GW amplitude birefringence. Following the general rules of eikonal approximation, we obtain
\begin{eqnarray}\label{EoMmag}
	\frac{dx^i}{dt} &=& -\frac{\partial D^\pm/\partial k_i}{\partial D^\pm/\partial\omega}  
	\approx  \{1 \pm i  (\alpha/\kappa) [2\partial_t\partial_z \phi + (\partial_t^2 \phi + \partial_z^2 \phi)] \} \delta^i_z\,,\nonumber\\
	\frac{dk_i}{dt} &=& \frac{\partial D^\pm /\partial x^i}{\partial D^\pm /\partial \omega} 
	\approx \mp i(\alpha \omega/2\kappa)  \left[ 2 \partial_t \partial_z\partial_i \phi + \partial_t^2 \partial_i \phi + \partial^2_z\partial_i \phi  \right]\,,\nonumber\\
	\frac{d\omega}{dt} &=& -\frac{\partial D^\pm /\partial t}{\partial D^\pm /\partial \omega} 
	\approx  \pm i(\alpha\omega/2\kappa)  \left[ 2 \partial_t^2 \partial_z \phi + \partial_t \partial_z^2 \phi + \partial_t^3 \phi  \right]\,,
\end{eqnarray}
where we have kept only the leading-order contributions and set $\omega = k$ to make our expressions simpler. For the GW path in which $x^i (t)$ is the function of time $t$, we can define the following total derivative of the scalar field
\begin{eqnarray}\label{PhiDot}
	\dot{\phi} \equiv \frac{d\phi}{dt} = \partial_t \phi + \partial_i \phi \frac{dx^i}{dt} \approx \partial_t \phi + \partial_z \phi\,, 
\end{eqnarray}
where we have used the leading-order GW velocity $dx^i/dt = \delta^i_z$. Hence, we can integrate the GW wavenumber and frequency over time $t$ in Eq.~\eqref{EoMmag} to obtain the following corrections
\begin{eqnarray}
	\Delta k_i &=& \mp i \alpha \omega \partial_i \dot{\phi}/(2\kappa)\,,  \quad
	\Delta \omega = \pm i \alpha \omega \partial_t \dot{\phi}/(2\kappa)\,.
\end{eqnarray}  
As a result, the GW phase $S$ varies as
\begin{eqnarray}
	\Delta S = -\int^{t_o}_{t_e} dt \Delta\omega + \int^{{\bf x}_o}_{{\bf x}_e} dx^i \Delta k_i 
	=  \mp i\alpha\omega (\dot{\phi}_o - \dot{\phi}_e)/(2\kappa) \,,
\end{eqnarray}
where the subscripts $o$ and $e$ represent the field values when GWs are observed and emitted, respectively. Therefore, the GW waveform would modify according to
\begin{eqnarray}\label{MagBi}
	h_{\rm R,L}  = h_{\rm R,L}^{\rm GR}   e^{i\Delta S} = h_{\rm R,L}^{\rm GR}  \exp\left( \pm \alpha \omega (\dot{\phi}_o- \dot{\phi}_e)/(2\kappa)\right),
\end{eqnarray} 
where $h_{\rm R,L}^{\rm GR} \equiv h^0_{\rm R,L} e^{iS_{\rm GR}}$ denotes the GW waveform predicted by GR. The above calculation clearly shows the GW amplitude birefringence.

\subsection{Observational Signatures}
Typically, the FDM mass is taken to be $m_\phi \sim {\cal O}(10^{-22}~{\rm eV})$, and  the corresponding de Brogile wavelength is of ${\cal O}(1\,{\rm kpc})$. Recent precise numerical simulations~\cite{Schive:2014dra,Schive:2014hza,Mocz:2017wlg} have shown that the clustering of such light FDM particles could form a core of flat density profile around the center of a MW-like galaxy, and suppress the formation of small structures, which could solve many problems faced by the cold DM~\cite{Robles:2018fur}. However, outside of the core, the FDM density transits into the conventional Navarro-Frenk-White (NFW) profile~\cite{Navarro:1996gj}
\begin{eqnarray}
	\rho_{\rm NFW} (r) ={\rho_0}\left[{\frac{r}{r_s} \left( 1+ \frac{r}{r_s} \right)^2}\right]^{-1}\,.
\end{eqnarray}
where $r_s$ and $\rho_0$ are two characteristic parameters. Thus, the local FDM field can be well estimated by~\cite{Khmelnitsky:2013lxt}
\begin{eqnarray}
	\phi(t,{\bf x}) = \frac{\sqrt{2\rho_{\rm NFW}}}{m_\phi} \cos\left(m_\phi t+\alpha({\bf x})\right)\,,
\end{eqnarray} 
which could reproduce the above NFW density distribution. Here $\alpha({\bf x})$ is a position-dependent random phase, which accounts for the incoherently fluctuating granular structures seen in simulations. In this subsection, we shall focus on the case with 

Now we concentrate on the GW propagation inside the MW, and postpone the discussion of the extra-galactic effects to the next section. Hence, we can take $\dot{\phi}_e$ and $\dot{\phi}_o$ in Eq.~\eqref{MagBi} to be the field values just outside our galaxy and around the Sun, respectively. However, the FDM density out of the MW can be estimated as $\rho_{\rm DM} = 0.265\rho_{\rm crit}$~\cite{Planck:2018vyg}, which is nearly six orders smaller than $\rho_\odot = 0.4\,{\rm GeV/cm^3}$ around the Solar system. Hence, the birefringence effect from $\dot{\phi}_e$ in Eq.~\eqref{MagBi} should be greatly suppressed and thus ignored. It turns out that the GW magnitude birefringence only depends on the local FDM property $\dot{\phi}_o$ near the Earth, giving us the following observed GW waveforms
\begin{eqnarray}\label{MagBiFDM}
	h^{\rm obs}_{\rm R,L} (f) = h^{\rm GR}_{\rm R,L} (f) \times \exp\left(\pm \frac{\kappa_A}{1\,{\rm Gpc}} \times \frac{f}{100\,{\rm Hz}} \right)\,,
\end{eqnarray}    
where $h^{\rm GR}_{\rm R,L}(f)$ are right- and left-handed the GW waveforms predicted by GR, while the opacity parameter is defined by
\begin{eqnarray}
	\kappa_A \equiv \pi\alpha \dot{\phi}_o/\kappa\,.
\end{eqnarray} 
From Eq.~\eqref{PhiDot}, $\dot{\phi}_o$ is the sum of the following local time and spatial derivatives of the FDM field near the Sun
\begin{eqnarray}
	 	 \partial_t \phi &=& \sqrt{2\rho_\odot} \sin\left(m_\phi t+\alpha_0\right)\,,\quad
	 	 \partial_z \phi \approx \cos\langle{\bf k}, {\bf r}\rangle \partial_r\phi \,,
\end{eqnarray}
where $R_\odot \simeq 8$\,kpc and $\alpha_0$ denote the radial galactic distance and the local field phase at the Solar system, respectively. The factor $\cos\langle{\bf k}, {\bf r}\rangle$ comes from the projection of the radial derivative of the FDM field $\partial_r\phi$  onto the GW propagating orientation ${\bf k}$ with
\begin{eqnarray}
	   	\partial_r\phi  = -\frac{\sqrt{{\rho_\odot}/{2}}}{m_\phi R_\odot}  \left(\frac{1+3R_\odot/r_s}{1+R_\odot/r_s}\right) \cos(m_\phi t +\alpha_0)\,.
\end{eqnarray}
However, for a FDM with $m_\phi \sim 10^{-22}\,{\rm eV}$, it is generally expected that the time variation of the field profile dominates over the birefringence due to $\partial_t\phi/\partial_r\phi \sim m_\phi R_\odot \sim {\cal O}(10^5) $. Therefore, the opacity parameter can be expressed as
\begin{eqnarray}\label{kAs}
	\kappa_A \simeq (\pi \alpha/\kappa) \sqrt{2\rho_\odot} \sin(m_\phi t + \alpha_0)\,.
\end{eqnarray}

Compared with the conventional signals studied in the literature, our predicted magnitude birefringence induced by the FDM shows several novel features. Firstly, it is obvious from Eq.~\eqref{MagBiFDM} that, due to its local nature, the birefringence factor is only a function of the GW frequency, without any dependence on the GW event distance, which is distinguished from the earlier results in \cite{Alexander:2007kv,Okounkova:2021xjv,Lagos:2024boe,Zhao:2019xmm,Jenks:2023pmk}. More significantly, our proposed birefringence in Eq.~\eqref{kAs} exhibits a remarkable time variation with the period directly reflecting the FDM mass. For example, if the mass is taken to be $10^{-22}\,{\rm eV}$, the oscillation period corresponds to 1.3 year, which can be viewed as a smoking gun of this FDM-generated GW birefringence.

\subsection{Effects from FDM Halo Inconherence}
We would like to mention that the above conclusion is derived under the assumption that GWs propagate over a single coherent region in which the axion field at different points oscillate with same phase. This is true when the axion is lighter than $10^{-24}$~eV, where its coherence length $l_c \sim 2\pi/(m_\phi v) \sim 40 \, {\rm kpc} \times (10^{-24}\,{\rm kpc}/m_\phi)$~\cite{Jung:2020aem,Tsutsui:2022zos,Tsutsui:2023jbk} can cover the whole galaxy for a typical DM velocity $v\sim 10^{-3}c$~\cite{Church:2018sro}. However, when the FDM mass increases to $m_\phi \sim 10^{-22}\,{\rm eV}$, the spatial coherence confines in a range of size $l_c \sim 2\pi/(m_\phi v) \sim 0.4\,\mathrm{kpc}$. Moreover, the duration of coherence can reach $t_c \sim 2\pi/(m_\phi v^2) \sim \mathcal{O}(10^6\, \mathrm{yrs})$~\cite{EPTA:2024gxu}, which is far longer than the time it takes for a GW to travel across the MW. Therefore, the Galaxy should be viewed to be composed of many FDM coherence patches. 
In each patch, the axion filed oscillates with the same pace in each patch. But the phases are randomly distributed among different patches over the FDM halo. Therefore, the GW waveform in Eq.~\eqref{MagBi} should be generalized to the multiple-patch case
\begin{eqnarray}
	h_{\rm R,L} = h_{\rm R,L}^{\rm GR} \exp(i\sum_n \Delta S_n) = h_{\rm R,L}^{\rm GR} \exp\left( \pm \alpha \omega \sum_n (\dot{\phi}_o^n - \dot{\phi}_e^n)/(2\kappa) \right)\,,
\end{eqnarray}
by assuming that the GW obeys the GR waveform at production. 
As pointed out in the previous subsection, the spatial derivative of the axion field profile is much less than its time derivative in most regions of the Galaxy. Thus, in what follows, we shall approximate $\dot{\phi}^n \approx \partial_t \phi^n = \sqrt{2\rho_{\rm NFW}} \sin(m_\phi t+\alpha_n)$ in the $n$-th coherent patch with its phase $\alpha_n$. We can further rearrange the total exponent into the following form
\begin{eqnarray}\label{MultiPatches}
	h_{\rm R,L} = h_{\rm R,L}^{\rm GR} \exp\left[ \pm \frac{\alpha \omega}{2\kappa} \sum_{n=1}^{N} \left(\dot{\phi}_o^{n-1} - \dot{\phi}_e^n\right) \pm \frac{\alpha\omega}{2\kappa} \left(\dot{\phi}_o^N -\dot{\phi}^0_e\right) \right]\,,
\end{eqnarray} 
where $\phi_e^0 = \phi_c$ is the field value at the outskirt of the MW while $\phi_o^N = \phi_\odot$ is the one at the Solar System. Now let us focus on one intermediate term $\dot{\phi}_o^{n-1} - \dot{\phi}_e^n$. Since the patches labeled by $n-1$ and $n$ are  adjacent to each other, the spacetime point for a GW exiting the region $n-1$ is the same one for its entrance to region $n$, {\it i.e.},  $(t_e^n, \mathbf{x}_e^n) = (t_o^{n-1}, \mathbf{x}_o^{n-1})$.  Thus, when a GW penetrates the boundary between these two regions, the only change of the field $\phi$ is the abrupt jump of the phase $\alpha_{n-1}$ and $\alpha_n$ while the amplitude of the axion oscillation keeps the same and is determined by the FDM density as $\sqrt{2\rho_{\rm NFW}(\mathbf{x}_e^n)}/m_\phi$. As a result, the intermediate term in Eq.~\eqref{MultiPatches} gives
\begin{eqnarray}
	\dot{\phi}^{n-1}_o - \dot{\phi}^n_e &=& \sqrt{2\rho_{\rm NFW}(\mathbf{x}_e^n)} \left[\sin(m_\phi t_e^{n} +\alpha_{n-1}) - \sin(m_\phi t_e^n + \alpha_{n})\right] \nonumber\\
	&=& 2\sqrt{2\rho_{\rm NFW}(\mathbf{x}_e^n)} \sin\left[ {(\alpha_{n-1}-\alpha_n)}/{2} \right] \cos\left[ m_\phi t_e^n + (\alpha_{n-1} + \alpha_n)/2 \right]\,.
\end{eqnarray}
Due to the random nature of the phase jump between the neighboring patches, its net effect on the GW propagation should be the average of jump factors, which tends to vanish as
\begin{eqnarray}
	\dot{\phi}^{n-1}_o - \dot{\phi}_e^n = \sqrt{2\rho_{\rm NFW} (\mathbf{x}^n_e)} \langle \sin\left[ (\alpha_{n-1}-\alpha_n)/2 \right] \rangle \langle \cos\left[ m_\phi t_e^n +(\alpha_{n-1} + \alpha_n)/2 \right] \rangle = 0\,.
\end{eqnarray}
Hence, all corrections of the FDM incoherence are expected to be zero in average, and finally we are left with only two terms either from the outskirt of the MW or at the observational site in the Solar System,
\begin{eqnarray}
	h_{\rm R,L} = h_{\rm R,L}^{\rm GR} \exp\left[ \pm {\alpha \omega (\dot{\phi}_\odot-\dot{\phi}_c)/(2\kappa)} \right]\,.
\end{eqnarray}
Hence, we have shown that the birefringece in Eqs.~\eqref{MagBiFDM} and \eqref{kAs} obtained in the coherent MW FDM halo is sitll valid in the presence of complicated spatial incoherence.

\section{Extra-galactic Contribution to Gravitational Wave Birefringence}\label{GWBiEx}
Beside the amplitude birefringence induced by the FDM in the MW, there is an additional contribution from the extra-galactic FDM field. One might worry that this contribution might affect or even govern the birefringence signal since it might be enhanced by the GW travel over an astrophysical distance. In order to investigate this important issue, we shall study the GW movement in the following cosmological FDM background~\cite{Kofman:1997yn,Machado:2018nqk}
\begin{eqnarray}\label{Phi0}
	\phi (t) = \phi_{0} \left( {a_0}/{a} \right)^{3/2} \cos(m_\phi t+ \alpha_c)\,,
\end{eqnarray}
where $\phi_0 = \sqrt{2\rho_{\rm DM}}/m_\phi$, $\alpha_c$ and $a_0$ are the FDM field amplitude, phase and present-day scale factor, respectively. Note that the field amplitude and phase should change at different spatial points, with the typical variation scale being the inverse of the de Broglie wavelength $m_\phi v$. For a small FDM velocity $v\ll 1$, such variations can be ignored compared with the dominant oscillation frequency $m_\phi$. Hence, we shall approximate the cosmological FDM profile to be homogeneous as in Eq.~\eqref{Phi0} to perform the following calculation. Also, we will set $a_0 =1$ for simplicity. The associated GW equation can be deduced from Eq.~\eqref{EoMGWt}
\begin{eqnarray}
	\square h_{\rm R,L} = \pm \frac{i\alpha}{\kappa a^2} \left[ \frac{1}{a^2} (\phi^{\prime\prime}-2 {\cal H}\phi^\prime)\partial_z h^\prime_{\rm R,L} - \phi^\prime \square \partial_z h_{\rm R,L}\right]\,,
\end{eqnarray}
which leads to the following dispersion relations for both circularly polarized modes 
\begin{eqnarray}
	\omega \equiv k + \Delta \omega^{\rm ex}  \approx k - i{\cal H} \omega/k \pm \frac{i\alpha\omega}{2\kappa a^2} (\phi^{\prime\prime}-2 {\cal H}\phi^\prime) \,,
\end{eqnarray}
up to the leading order in the eikonal approximation. 

For the FDM scalar with $m_\phi \sim 10^{-22}~{\rm eV}$, the mass scale is much larger than the cosmological expansion rate characterized by ${\cal H}$. Thus, we expect that the time integration in the  birefringence factor $e^{i\Delta S} = e^{-i\int d\eta \Delta \omega^{\rm ex}}$ should be dominated by the rapidly oscillating term in $\phi^{\prime\prime}$. As a result, the phase correction generated by the extra-galactic FDM background is given by
\begin{eqnarray}
  \Delta S \approx \pm i\left(\frac{\alpha \omega}{2\kappa}\right)  \sqrt{2\rho_{\rm DM}} \sin(m_\phi t +\alpha_c)\,,
\end{eqnarray}
in which we have taken the small redshift limit with the scale factor being $a \approx 1$.  Hence, the FDM outside of the MW would give the following amplitude birefringence 
\begin{eqnarray}
	h^{\rm ex}_{\rm R,L}(f) = h_{\rm R,L}^{0} (f) \exp\left(\pm \frac{\kappa_A^{\rm ex}}{1~{\rm Gpc}} \times \frac{f}{100~{\rm Hz}}\right)\,,
\end{eqnarray}
where $\kappa_A^{\rm ex} \equiv \alpha\pi \sqrt{2\rho_{\rm DM}} \sin(m_\phi t +\alpha_c) /\kappa$. In comparison with Eqs.~\eqref{MagBiFDM} and \eqref{kAs}, it is obvious that the birefringence effect is overwhelmed by the galactic FDM component due to its enhanced DM density in the MW, which is evident by $|\kappa_A/\kappa_A^{\rm ex}|  \sim \sqrt{\rho_\odot/\rho_{\rm DM}}\sim {\cal O}(10^3)$.  

\section{Conclusion and Discussion}\label{conclusion}
The FDM is a promising DM candidate which can possibly solve many problems faced in the sub-galactic scale. If such a FDM can be identified as an axion-like particle with an additional gravitational CS coupling, GWs are expected to show the parity-violating birefringence phenomena when propagating in the nontrivial FDM background. Especially, provided the complicated granular structures displayed in recent simulations, we are led to considering the GW propagation in a general spacetime-dependent FDM field profile. By using the famous eikonal approximation, we find that GWs do not exhibit any velocity birefringence in the CS gravity, no matter if there is a FDM background field. However, the inclusion of the imaginary part in the GW dispersion relations gives rise to the amplitude birefringence, {\it i.e.}, one circular polarization is enhanced whereas the other suppressed. Due to the local nature of this galactic birefringence, the obtained effect only depends on the GW frequency without any reliance on the GW event distance. More remarkably, such amplitude modifications of the left- and right-handed polarizations oscillate in time with the period controlled by the FDM scalar mass. {We have taken into account the spatial incoherence of the FDM field in the galactic halo, finding that this effect tends to vanish in average due to the random distribution of the field phase over the galaxy.} Finally, the extra-galactic FDM-induced contribution to the GW birefringence has also been considered, which is shown to be subdominant owing to the suppression from the cosmological DM density. 

Currently, the existing LVK GW data already allows us to constrain the predicted FDM-induced amplitude birefringence.  By including the associated birefringence factor in Eq.~\eqref{MagBiFDM} into the left- and right-handed GW components, one can compare the obtained waveform model against the whole set of compact binary events~\cite{Okounkova:2021xjv,Zhao:2019xmm,Ng:2023jjt} or the binary neutron star merger GW170817 with a multi-messenger approach~\cite{Lagos:2024boe}. 
Nevertheless, as emphasized before, the birefringence factor produced by the FDM is only a function of the GW frequency and does not rely on the GW propagation distance, which is a novel signature and requires a new fit to the LVK datasets. We can estimate the bound on $\kappa_A$ to be of  ${\cal O}(0.01~{\rm Gpc})$ based on previous investigations in Refs.~\cite{Okounkova:2021xjv,Zhao:2019xmm,Lagos:2024boe,Ng:2023jjt}. More significantly, if the measured $\kappa_A$'s for GW events over a timescale of a few years clearly show the periodic time dependence, this would further support the FDM origin of the GW birefringence. However, a detailed discussion of the data-analysis issue is beyond the scope of the present paper, and we would like to postpone this exploration to a future work.
 
\begin{acknowledgements}
\noindent DH would like to thank Prof.~Kazuya Koyama for his warm hospitality at the Institute of Cosmology and Gravitation, University of Portsmouth, where part of this work was carried out. DH is also grateful to Prof. Ian Harry and Prof. Tessa Baker for enlightening discussions. This work is supported in part by the National Key Research and Development Program of China under Grant No.~2024YFC2207204 and No.~2021YFC2203003 , and the China Scholarship Council under Grant No.~202310740003.
\end{acknowledgements}

\appendix

\bibliography{GWBi}

\end{document}